\documentclass[reprint,showpacs,showkeys,twocolumn,10pt]{revtex4-1}
\usepackage{graphicx}
\usepackage{times}
\usepackage{amsmath}

\newcommand{\Heb}{H_{eb}}
\newcommand{\Hc}{H_{c}}
\newcommand{\Jfm}{J_{FM}}
\newcommand{\Jaf}{J_{AF}}
\newcommand{\Jint}{J_{int}}

\newcommand{\Nc}{N_c}
\newcommand{\Nu}{N_{u}}
\newcommand{\Rc}{R_c}
\newcommand{\Lc}{L_c}
\newcommand{\tsh}{t_{sh}}

\begin{document}
\title[Exchange-bias in magnetic nanowires]
{Monte Carlo study of the exchange bias effect in Co/CoO core-shell nanowires}
\author{A.~Patsopoulos$^1$}
\author{D.~Kechrakos$^{2}$}
\email{dkehrakos@aspete.gr}
\affiliation{$^1$Department of Physics, National and Kapodistrian University of Athens, Athens, GR-15784}
\affiliation{$^2$Department of Education, School of Pedagogical and Technological Education, Athens, GR-14131 }
\date{\today}

\begin{abstract}

We study the magnetic properties of cylindrical ferromagnetic core - antiferromagnetic shell nanowires using Monte Carlo simulations and a classical Heisenberg Hamiltonian in order to elucidate the impact of the oxidized shell on the magnetic properties and the magnetization reversal mechanism. 
We find that the coupling to the antiferromagnetic shell leads to  suppression of the coercivity and emergence of a weak exchange bias effect. 
Comparison of the magnetization reversal mechanism in the bare and the surface-oxidized nanowire reveals that the
domain wall propagation and annihilation remains the dominant reversal mechanism in surface oxidized nanowires as in their ferromagnetic counterparts. 
However, the interface exchange coupling introduces a secondary reversal mechanism activated in the central part of the wire with characteristics of coherent rotation, which acts in synergy to wall propagation leading to enhancement of the wall mobility. 
This effect is more pronounced in nanowires with large exchange bias values and is attributed to the uncompensated interface moments that act as nucleation centers for magnetization reversal. 
Our results are in good agreement with recent measurements in Co and Co/CoO nanowires.
\end{abstract}
\maketitle
\section{Introduction}
Elongated magnetic nanostructures, such as nanorods and nanowires are characterized by enhanced anisotropy due to their shape
and hold promises for major advances in different areas of modern technology ranging from magnetic recording\cite{sel01} and spintronics\cite{sel01} to biomedicine\cite{bau04,tar06}.
A new perspective in magnetic memory devices has also emerged\cite{par08}, stimulated by the manifested feasibility to manipulate the domain wall motion in these quasi one-dimensional nanostructures and paved new paths for information storage and spintronics applications\cite{hrk11}.
A central aim of fundamental research related to magnetic nanostructures remains to reveal the various factors that govern the magnetization reversal mechanism.
The well established coherent rotation model of Stoner and Wolfarth \cite{tan08}  describes accurately the magnetization dynamics of ferromagnetic nanostructures with diameter up to a few nanometers and has been experimentally verified on individual nanoparticles\cite{wen97}. 
In ferromagnetic nanowires the magnetization reversal process is more complex and consists of three steps that include nucleation, propagation and annihilation of domain walls\cite{fer99,sel01,thi06}.
The situation gets more complicated as a the diameter of a FM nanowire was shown\cite{hin00a,wie04} to control the character of the domain walls and drives a transition from a transverse domain wall to a vortex domain wall as the diameter increases beyond the exchange length.
In a parallel effort to develop magnetic materials with desired properties, the exchange bias effect\cite{mei56,mei57,nog99} has long been recognized as a means to tailor the hysteresis characteristics of nanostructured magnetic materials\cite{nog05,igl08} and is temporarily implemented in spin-valves and magnetic tunnel junctions that are constituent elements of spintronics devices.
In contrast to the large amount of research devoted to studies of the exchange-bias effect in coupled ferromagnetic/antiferomagnetic (FM/AF) bilayers\cite{nog99} and nanoparticles with core-shell morphology\cite{igl08}, 
the field of exchange coupled FM/AF nanowires with core-shell morphology remains relatively unexplored.
Experimental studies, demonstrated exchange bias behavior in cylindrical permalloy nanowires\cite{buc15}, oxidized Co nanowires\cite{mau09} and  nanotubes\cite{pro13} with characteristic accompanying effects, such as loop shift and training effect, previously reported for exchange-biased nanoparticles\cite{igl08}.
The competition between shape anisotropy, cooling field and applied field directions was shown to lead to a variety of novel properties of core-shell Co/CoO nanowires, such as, tailor-made magnetic response\cite{tri10} and high-field irreversibility accompanied by cooling-field dependent magnetization\cite{sal16}.
Furthermore, Maurer \textit{et al}\cite{mau09} compared the hysteresis properties of cylindrical Co and Co/CoO nanowires and demonstrated the suppression of the coercive field due to surface oxidation as well as an anomalous temperature dependence, which they attributed to the thermal fluctuations of the oxide shell. This study provides additional evidence that the exchange-bias effect drastically shapes the magnetization reversal mechanism of magnetic nanowires.
Numerical studies of hysteresis properties of FM nanowires\cite{her02,for02} are commonly based on the micromagnetic theory\cite{aha96}. 
Micromagnetic studies of the exchange-bias effect in FM-AF structures, usually adopt a "frozen-field" approximation \cite{mau09, spi15}, in the description of the AF component, that completely neglects the thermal fluctuations of the AF.
Instead, an atomistic approach\cite{eva14} accounts correctly for the dynamics of both phases  in a coupled FM-AF system, but currently much smaller systems can be handled, due to enormous requirements in computer resources when even samples with typical length of $\approx 1\mu m$ have to be modeled.
The Monte Carlo method has proved to be a versatile and reliable approach to study thermal effects in the magnetization dynamics of complex nanostructures\cite{igl08} as well as isolated FM nanowires.
Hinzke and Nowak \cite{hin00a} using the Monte Carlo method  demonstrated the transition form a transverse to a vortex domain wall in a FM nanowire. The same authors\cite{hin00a,hin00b} provided numerical evidence for the equivalence of the Monte Carlo method to the full dynamical approach based on the stochastic Landau-Lifshitz-Gilbert equations in the high damping limit.
More recently, Allende \textit{et al}\cite{all08} demonstrated the nucleation and propagation of transverse domain walls in FM nanowires and found a complex behavior in the domain wall propagation that depends on the strength and orientation of the applied field. 

In the present work, we use the Monte Carlo method to treat the magnetic thermal fluctuations in modeling the hysteresis behavior of cylindrical nanowires composed of a FM core and an AF shell, thus going beyond the "frozen-field" approximation to the exchange-bias effect. 
We examine on a microscopic level the interplay between exchange biasing and domain wall propagation during field-driven magnetization reversal.
We find that exchange biasing increases the domain wall velocity and mobility and modifies the magnetization reversal mechanism. The uncompensated spins at the FM-AF interface promote a coherent reversal mechanism of the core magnetization, which is absent in bare FM nanowires.

\section{Modeling and Simulation Method}
\begin{figure}[htb!]
\centering
\includegraphics[width=0.950\linewidth]{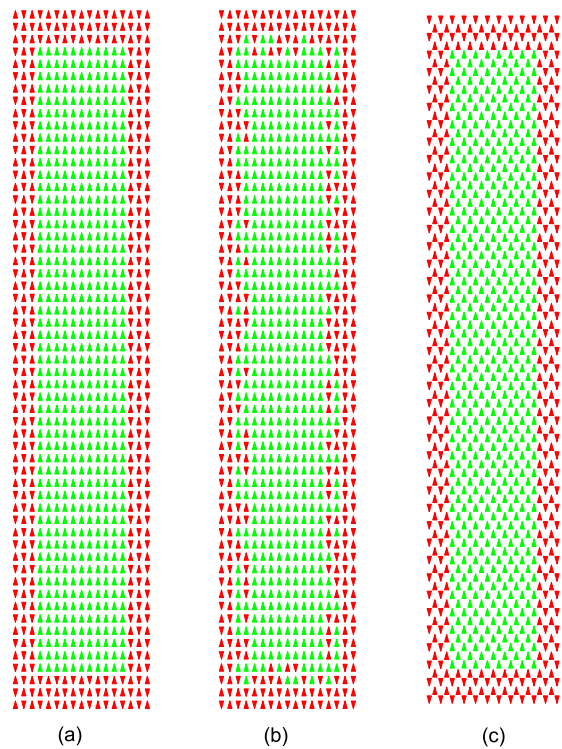}
\caption{ 
Cutting planes for different structural models of cylindrical core-shell nanowires with aspect ratio 5:1. 
The color code distinguishes the FM core sites (light green) from the AF shell sites (red). Arrowheads show the moment distribution at the FC state. 
(a) CS[001], 
(b) CS[001]-rough with 50\% site intermixing, and 
(c) CS[011].
The shell interface layer in case (c), consists of AF sites belonging to the same magnetic sublattice (up-spin) leading to a highly uncompensated interface.
Parameters: $R_c=5a$, $L_c=50a$ and $\tsh=3a$.
}
\label{fig:sites}
\end{figure}

Nanowires are generated by cutting a cylinder with radius $R$ and length $L$ from a simple cubic (sc) lattice with lattice constant $a$. 
For the core-shell morphology we define an internal homoaxial cylinder with radius $R_c = R-t$ and length $L_c=L-t$ , where $t$ is the shell thickness (Fig.\ref{fig:sites}).
The microstructural details at the interface region are of crucial importance in studies of exchange bias in FM-AF coupled systems, because they determine the number of uncompensated interface moments\cite{eva11,dim15}.
The degree of uncompensation is obtained by counting the up and down spins in the AF shell, $\Nu=N_+-N_-$. Obviously, any modifications in the interface microstructure will modify the number of uncompensated spins.
In the structural models studied here, which are sketched in Fig.\ref{fig:sites}, we vary the degree of compensation at the FM-AF interface in two ways: 
(i) By growing the nanowire along the [011] direction of the sc lattice. This choice leads to a highly uncompensated interface, because consecutive AF sites along the z-axis belong to the same sublattice. 
(ii) By introducing atomic-scale roughness through random intermixing of the FM and AF sites in the interface region. This procedure replaces the clean  interface layers $A|B$ by an $A_{1-x}B_x|B_{1-x}A_x$ random alloy. Below we consider the case $x=0.5$ in order to have the maximum effect of spin uncompensation due to interface intermixing. 
Structural parameters for nanowires with aspect ratio $L_c/D_c=5$ are summarized in Table~\ref{table:struct}.
\begin{table}    
  \caption{
  Structural parameters of nanowires with $\Rc=5a$, $\Lc=50a$ and $\tsh=3a$ (core-shell only). $N=~$total nr of sites in a nanowire, $\Nc=~$nr of core (FM) sites and $\Nu=~$nr of uncompensated (AF) sites.
  }
  \begin{ruledtabular}
  \begin{tabular}{l l l l}     
  System        &$N$    &$\Nc$  &$\Nu$  \\
  \hline
  FM[001]       & 4131  &4131          & -     \\
  CS[001]       &11229  &4131          &  2    \\
  CS[001]-rough &11229  &4168$^{(1)}$  & 27$^{(1)}$\\	
  FM[011]       & 3867  &3867          & -     \\
  CS[011]       &11021  &3867          & 66    \\
  \end{tabular}
  \end{ruledtabular}
  $^{(1)}$ Average value over $25$ realizations of interface disorder
  \label{table:struct}
\end{table}

The total energy of the magnetic system reads 
\begin{equation}
  E = \sum_i E_i   
  \label{eq:totenergy}
\end{equation}
with the single-site energy
\begin{eqnarray}
  E_i = 
  -\frac{1}{2} \widehat{S}_i \cdot \sum_{<j>} J_{ij}  \widehat{S}_j 
  -K_i S_{i,z}^{~2}  \nonumber \\
  -H S_{i,z}
  -\frac{1}{2} g \widehat{S}_i \cdot \sum_{j}  \textit{\textbf{D}}_{ij} \cdot \widehat{S}_j. 
  \label{eq:energy}
\end{eqnarray}
In Eq.~(\ref{eq:energy}) and further on, hats indicate unit vectors and bold symbols indicate $3\times 3$ matrices in Cartesian coordinates. 
The factor $1/2$ in front of the first and fourth term of Eq.~(\ref{eq:energy}) accounts for the double-counting in the calculation of the total energy, Eq.~(\ref{eq:totenergy}).
The first term in Eq.~(\ref{eq:energy}) is the exchange energy between first nearest neighbor (1nn) sites. 
The exchange constant  $J_{ij}$ takes the values $J_{FM}$, $J_{AF}$ and $J_{int}$ depending on whether sites $i$ and $j$ belong to the FM region, the AF region or the interface region, respectively. 
The latter extends over the sites of the core (shell) having exchange bonds with sites in the shell (core). 
For 1nn exchange couplings, the interface region has width $2a$ and consists of the core interface layer and the shell interface layer.
The second term in Eq.~(\ref{eq:energy}) is the uniaxial anisotropy energy with the easy axis taken along the cylinder axis (z-axis). 
The anisotropy constant takes the values $K_{FM}$ and $K_{AF}$ depending on the location of site $i$. 
The third term in Eq.~(\ref{eq:energy}) is the Zeeman energy due to an external field of strength $H$ and the last term is the dipolar energy with strength $g$. 
The dipolar matrix $\textit{\textbf{D}}$ is defined as 
\begin{equation}
D_{ij}^{\alpha\beta} = 
( 3 r_{ij}^{\alpha} r_{ij}^{\beta} - {\delta}_{\alpha\beta} )  / (R_{ij}/a)^3
\label{eq:dmatrix}
\end{equation}
with $R_{ij}$ the distance between sites $i,j$ and $\widehat{r}_{ij}$ is the unit vector along the direction from site $i$ to site $j$. Indices $\alpha, \beta$ denote Cartesian coordinates.
The exact computation of the dipolar field, namely the sum in the last term of Eq.~(\ref{eq:energy}), is a computationally demanding task due to the infinite range of dipolar interactions. 
To tackle this problem we decompose the dipolar field into a near-field and a far-field component and implement for the latter a mean-field approximation\cite{rus01}, extended to the case of a two-phase system with free boundaries.
In particular, we write
\begin{eqnarray}			
  \sum_{j}  \textit{\textbf{D}}_{ij} \cdot \widehat{S}_j = 
  \sum_{j,~r_{ij} \leq r_0} \textit{\textbf{D}}_{ij} \cdot \widehat{S}_j + 
  \sum_{j,~r_{ij}    > r_0} \textit{\textbf{D}}_{ij} \cdot \widehat{S}_j 
  \label{eq:ddsplit}
\end{eqnarray}			
where $r_0$ is a cutoff distance defining the range of the near field.
We approximate the second term on the right hand side of the above expression as 
\begin{eqnarray}
  \sum_{j, ~r_{ij} > r_0} \textit{\textbf{D}}_{ij} \cdot \widehat{S}_j \approx~ \textit{\textbf{d}}_{i}^{FM} \cdot \langle\widehat{S}\rangle_{FM} +~
  \textit{\textbf{d}}_{i}^{AF} \cdot \langle\widehat{S}\rangle_{AF}
  \label{eq:farfield}
\end{eqnarray}
where
\begin{equation}
  \textit{\textbf{d}}_{i}^{FM(AF)}= 
  \sum_{\substack{j \in FM (AF) \\ r_{ij}>r_0}} \textit{\textbf{D}}_{ij}
  \label{eq:demag}
\end{equation}
is the demagnetization matrix on site $i$ and 
$\langle \widehat{S} \rangle_{FM(AF)}$ is the average spin over the FM (AF) region of the core-shell nanowire. 
Notice that the mean-field approximation adopted in Eq.~(\ref{eq:farfield}), has a \textit{local} character to account for the different environment of each site in a system with free and internal boundaries. This approximation leads to a site-dependent demagnetization matrix. 
Furthermore, the two-phase character of the core-shell system is preserved as indicated by the distinct spin averages over the two phases of the composite system in Eq.~(\ref{eq:farfield}).

In order to observe shifted hysteresis loops due to exchange bias, the CS nanowire has to be field-cooled (FC) from a high temperature $(T>>T_c)$ to a low temperature $(T<<T_N)$ in the presence of a cooling field that is well below the saturation field of the AF phase $(H_{cool} << H_{sat})$. 
At the end of the FC process, the FM spins are aligned along the external field, while the shell interface spins are frozen parallel or antiparallel to the core interface spins. 
To avoid the time-consuming simulation of the FC procedure, we adopt N\'{e}el's two-sublattice model for the AF and approximate the FC state of the shell by the state which minimizes the total energy\cite{eva11,dim15}. 
Using this state as the initial spin configuration of the system, we sweep the external field $( -H_{cool} \le H \le + H_{cool} )$ at a constant rate to obtain the isothermal hysteresis loop.
The effective coercivity of the system is then defined as 
$\Hc= |H_{c1} -H_{c2}|/2$
and the exchange bias field as
$\Heb= |H_{c1} + H_{c2}|/2$, where $H_{c1}$ and $H_{c2}$ are the left (negative) and right (positive) coercive fields of the FC system.

For the simulation of the hysteresis loops we use the Metropolis Monte Carlo algorithm with single spin updates. 
Trial spin moves were confined in a cone of angle $\sim3^o$ around the initial spin direction, which resulted in an average acceptance ratio 40-60\% of the attempted moves. 
Thermal average were taken over 
$N_{rs}=10$ independent relaxation sequences, each composed of 
$M_0=0.5 \times  10^{4}$ Monte Carlo steps per spin (MCSS) for thermalization and  
$M= 10^{4}$ MCSS for measurements.
Measurements are performed every 
$\tau=10$ MCSS to minimize correlations between sampling points. 
The field sweep rate is kept constant at 
$r_H=10^{-5} \Jfm/$MCSS, to exclude the  variation of results with sampling time.
Finally, when intermixing at the FM-AF interface is considered, the results are averaged over 
$N_c=25$ configurations of the interface randomness. 
In the computation of dipolar fields we use a truncation radius $r_0=3a$, which introduces an estimated error of less than $1\%$ in the total energy of a uniformly magnetized FM nanowire. 

In our simulations we scaled all energy parameters entering Eq.(\ref{eq:energy}) with $J_{FM}$, which for numerical convenience was given the arbitrary value $J_{FM}=10$.
Then, we define 
$J_{AF}$ = -0.5$J_{FM}$, 
$J_{int}$ = -0.5$J_{FM}$, 
$K_{FM}$ = 0.1$J_{FM}$, 
$K_{AF}$ = 1.0$J_{FM}$ and $g=0.05J_{FM}$.
These parameters capture the main features of the Co/CoO exchange coupled system as previous studies on magnetic nanoparticles have shown\cite{igl08,dim15} and are similar to discretized material parameters used in micromagnetic studies\cite{wys10,tos16,comm1b}. 
As for the $J_{int}$ values, due to lack of clear experimental data we adopted a value previously used, namely $J_{int} \approx J_{AF}$ \cite{igl08,dim15}. 
For simplicity, we assume that the magnetic moments of the ferromagnetic and antiferromagnetic sites  are equal ($\mu_F=\mu_{AF}$), thus the parameters $H$ and $g$ become independent of site.
A large value of shell anisotropy $(K_{AF} \simeq J_{AF})$  is used as in previous Monte Carlo simulations of exchange-biased nanostructures\cite{igl08,dim15}, which is justified by the fact that for thin shells there is an important contribution to the shell anisotropy that arises from the low symmetry near the free surface of the nanostructure\cite{dim15}. 
The presence of random anisotropy in the outermost shell layer would be appropriate for polycrystalline shells. However we restrict ourselves here to high quality shells, as those reported in recent experiments on Co/CoO nanowires\cite{mau09}
\section{Numerical results}
\subsection{Hysteresis loops}
\begin{figure} [htb!]
	\centering
	\includegraphics[width=0.95\linewidth]{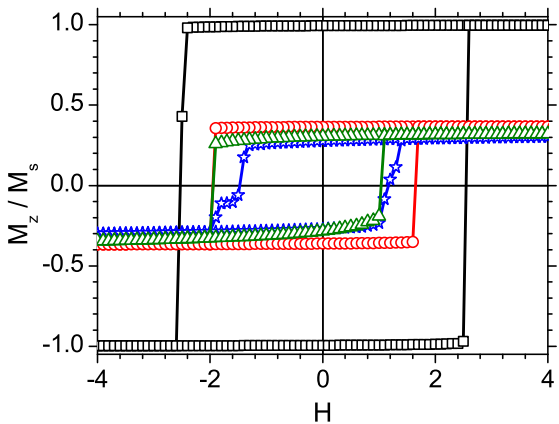}
	\caption{ 
		Isothermal ($T=0.01J_{FM}/k_B$) hysteresis loops of the total magnetization for a FM and different FM-AF nanowires. In all cases the FM region has $\Rc=5a$ and $\Lc=50a$ and the shell thickness $\tsh=3a$. (Black) squares : FM[001], (red) circles : CS[001], (blue) stars : CS[001]-rough and (green) triangles: CS[011] nanowires.
	}
	\label{fig:loops}
\end{figure}

We examine first the macroscopic magnetic behavior of nanowires, by calculating their hysteresis loops and their characteristic fields, namely $\Hc$ and $\Heb$. 
In Fig.~\ref{fig:loops} we compare the low-temperature isothermal hysteresis loops of bare FM and core-shell FM-AF nanowires with different interface microstructure (CS[001], rough-CS[001], CS[011]) but the same size of the FM core.
The main features of these data are  a substantial reduction of the coercivity relative to the bare FM nanowire and an emergence of a weak exchange bias field, leading eventually to encapsulation of the CS loop inside the FM loop.  
The low values of $\Heb$ of AF-coated nanowires is due to the fact that cylindrical interfaces are highly compensated compared to their spherical counterparts as a result of the atomic planes stacking along the nanowire axis. 
This geometrical effect is more pronounced in nanowires grown along the [001] axis, while interface roughness or growth of the nanowire  along the [011] axis increases the number uncompensated moments (Table~\ref{table:struct}).
In the first case, roughness increases the number of uncompensated spins in a statistical manner, while in the second case growth of the nanowire along the [011] direction generates an ordered uncompensated interface, because successive AF sites along the nanowire axis belong to the same AF sublattice (Fig.~\ref{fig:sites}). 
Despite the similar outcome of these two structural factors, they lead to distinct features in the overall shape of the hysteresis loop.
Interface roughness has a clear loop shearing effect which is absent in the case of a nanowire with a lower symmetry axis. 
The loop shearing is the outcome of the averaging process over an ensemble of nanowires with different disorder realizations, that is characterized by a distribution of coercive fields, and has been previously reported in studies of core-shell magnetic nanoparticles with rough interfaces\cite{dim15,eva11}.
\begin{figure} [htb!]
	\centering
	\includegraphics[width=0.95\linewidth]{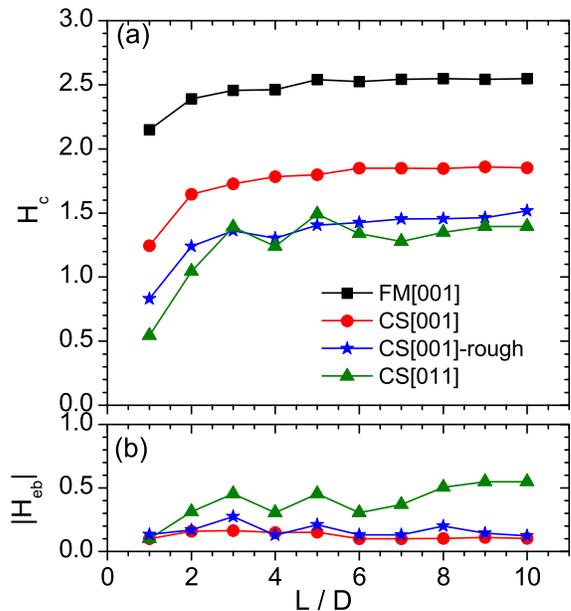}
	\caption{
		Dependence of (a) coercivity and (b) exchange bias field on aspect ratio for nanowires with constant core radius $\Rc=5a$ and shell thickness $\tsh=3a$. Temperature $T=0.01 \Jfm /k_B$.
	}
	\label{fig:aspratio}
\end{figure}

The characteristic fields $\Hc$ and $\Heb$ are expected to show a dependence on the length of the cylindrical nanostructure (Fig.~\ref{fig:aspratio}). 
For the FM[001] nanowire, $\Hc$ shows an initial increase with nanowire length due to increasing shape anisotropy and eventually it reaches a saturation value at $L_c/D_c \approx 4$.
This result is in qualitative agreement with recent measurements in Co nanorods where coercivity saturation was measured at approximately the same aspect ratio values\cite{pou15}.
The same dependence of $\Hc$ on nanowire length is seen when for the CS[001] nanowires, however the saturation value of $\Hc$ is lowered due to exchange coupling to the AF shell.
Oxide-coated nanowires with interface roughness (rough-CS[001]) or  lower symmetry axes (CS[011]), exhibit larger suppression of $\Hc$. This points to a coercivity reduction mechanism driven by the uncompensated moments, the details of which are discussed below. 
On the other hand, $\Heb$ is only weakly dependent on cylinder length (Fig.~\ref{fig:aspratio}), because it arises from a predominantly interface effect and as such it is expected to scale with the surface-to-volume ratio of the nanostructure. 
For a cylindrical shape this ratio is proportional to the inverse core radius and therefore independent of cylinder length.
\begin{table}    
\caption{
Characteristic fields of nanowires with $L_c/D_c=9$ and $\tsh=3a$ (core-shell only). 
}
\begin{ruledtabular}
\begin{tabular}{l l l l l }     
System          &$\Hc$   &$\Heb$ &$\Hc/H_{c,FM}$  &$\Heb/\Hc$ \\
\hline
FM[001]         &2.54    &0.00   &1.00            &0.00       \\
CS[001]         &1.86    &0.11   &0.73            &0.06       \\
CS[001]-rough   &1.46    &0.14   &0.57            &0.10       \\
CS[011]         &1.39    &0.55   &0.55            &0.40       \\
Co/CoO$^{(1)}$  & -      & -     &0.55            &0.20       \\
\end{tabular}
\end{ruledtabular}
$^{(1)}$ Experimental data from Ref.\cite{mau09}\\
\label{table:hchb}
\end{table}

As a final remark, we compare our simulation results for the low temperature hysteresis loops to the measurements of Maurer et al \cite{mau09} on Co and Co/CoO nanowires.
The calculated large suppression of $\Hc$ and the weak exchange bias effect resulting in encapsulation of the CS loop inside the FM loop,  as shown in Fig.~\ref{fig:loops}, have been also observed experimentally\cite{mau09}.
Since a direct comparison of the calculated values of $\Hc$ and $\Heb$ and the experimental is not possible within our model, we restrict ourselves to a comparison of relevant ratios.
In particular, we quantify the suppression of coercivity due to interface exchange coupling by the ratio $\Hc/H_{c,FM}$ and the weakness of the bias field by the ratio $\Heb/\Heb$.
In Table~\ref{table:hchb} we summarize the calculated data for nanowires with aspect ratio similar to the experimental ones\cite{mau09}.
As seen there, when interface uncompensation is included in the structural model, either through interface roughness or symmetry of the interface, our simulation data are in good agreement to the measurements. This is reasonable, as atomic scale roughness and misorientation of core and shell lattice structures is inherent to the chemical preparation method\cite{mau09} of nanowire samples.
\subsection{Magnetization reversal mechanism}
Next, we discuss the microscopic magnetization reversal mechanism of bare FM and AF-coated nanowires and underline the differences introduced by the core-shell morphology. 
It is well known that in FM nanostructures with all three dimensions below the exchange length, as in nanoparticles and short nanorods, the magnetization reversal proceeds by coherent rotation\cite{ber98}.
On the contrary, in FM nanowires that are characterized by large aspect ratio, the magnetization reversal is realized by propagation and annihilation of a domain wall pair that nucleate at the two free ends of the nanowire \cite{fer99,hin00a,thi02}.
The transition from the regime of coherent rotation for short nanorods to the regime of domain wall propagation in long nanowires manifests itself macroscopically in the increasing values of $\Hc$ with aspect ratio, as in Fig.~\ref{fig:aspratio}.
Saturation of the coercivity values indicates the establishment of the domain wall propagation mechanism. 
According to our results in Fig.~\ref{fig:aspratio}, domain wall propagation should be observed in nanowires with aspect ratios $L_c/D_c \gtrsim 4$.
Thus, in the rest of this paper, we study magnetization dynamics of nanowires with $L_c/D_c =5$.

\begin{figure} [htb!]
\centering
\includegraphics[width=0.95\linewidth]{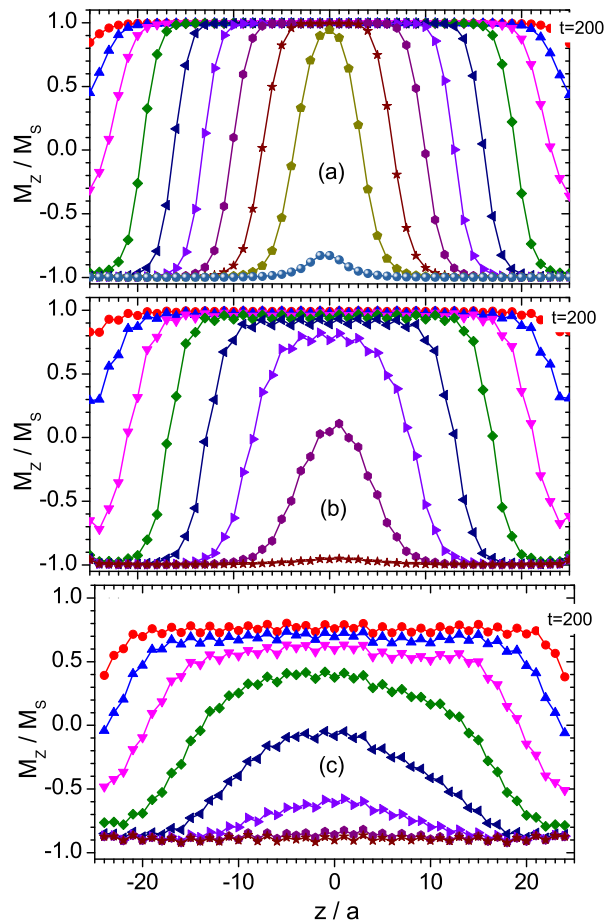}
\caption{Time-evolution of magnetization profile under application of a reverse field $H=-0.5\Jfm$. Snapshots are taken every $\Delta$t=200 MCS starting at $t_0$=200 MCS (uppermost curve).
(a) FM[001] nanowire,
(b) CS[001] nanowire, and 
(c) CS[001]-rough nanowire. 
Parameters $R_c=5a, \tsh=3a$, $L_c=50a$ and $T=0.01\Jfm$.
}
\label{fig:mprof}
\end{figure}

\begin{figure} [htb!]
\centering
\includegraphics[width=0.95\linewidth]{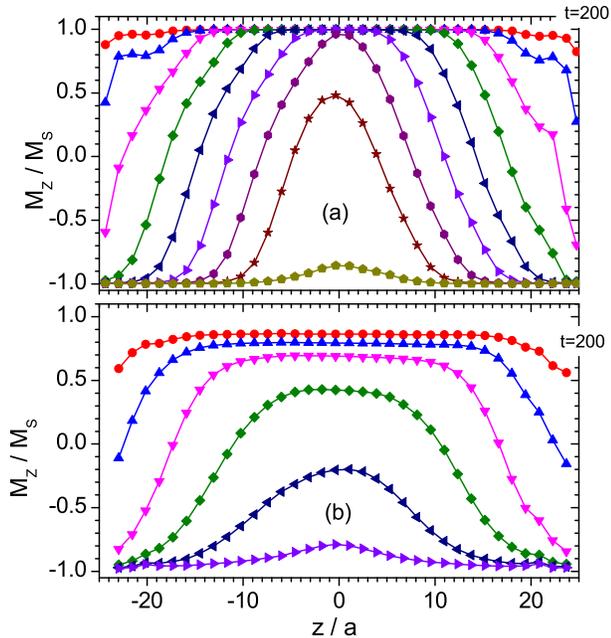}
\caption{Time-evolution of magnetization profile under application of a reverse field $H=-0.5\Jfm$. Snapshots are taken every $\Delta$t=200 MCS starting at $t_0$=200 MCS (uppermost curve).
(a) FM[011] nanowire, and 
(b) CS[011] nanowire. 
Parameters $R_c=5a, \tsh=3a$, $L_c=50a$ and $T=0.01\Jfm$.
}
\label{fig:mprof_rot}
\end{figure}

To study the reversal dynamics, we start our simulation with a nanowire in the FC state, we apply a reverse field and record the time evolution of the magnetization profile $M_z(z)$, which is defined as the decomposition of the total core magnetization into contributions from atomic planes normal to the nanowire axis,
\begin{equation*}
  M_z(z)=\sum_{i\in FM}  S_{i,z} \cdot \delta(z_i-z) / 
  \sum_{i\in FM}\delta(z_i-z).
\end{equation*}
Results for the time-evolution of the magnetization profile are shown in Fig.~\ref{fig:mprof} and Fig.~\ref{fig:mprof_rot}, where the three-stage magnetization reversal mechanism, namely nucleation-propagation-annihilation is clearly seen. 
From examination of the in-plane distribution of the magnetization in the region of the walls we deduced that in all cases studied here, the nanowires support transverse domain walls\cite{hin00a,wie04}.
This observation was further supported by calculation of the parameter $M_w$, defined as
\begin{equation}
  M_w(z) =\sum_{i\in FM} 
  [ \widehat{r}_i \times  \widehat{S}_i ]_z \delta(z_i-z) / 
  \sum_{i\in FM}\delta(z_i-z) 
\end{equation}
that measures the degree of magnetization winding \cite{ter16}.
Our calculations showed $M_w(z)\approx 0$ at all time steps during the magnetization reversal. 
Therefore, the longitudinal component $M_z(z)$ contains all the information about the domain wall dynamics. 

\begin{figure}[htb!]
\centering
\includegraphics[width=0.95\linewidth]{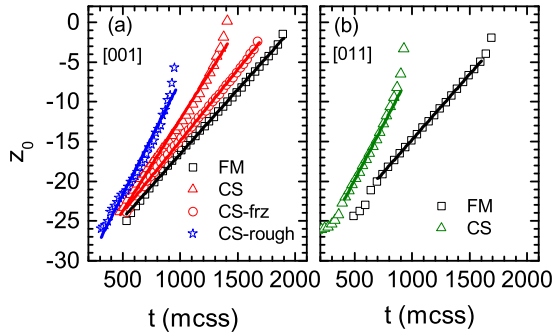}
\caption{ 
Time-evolution of domain wall position.
(a) FM, CS and CS-rough nanowires along [001]. Freezing the shell spins during reversal (CS$_{frz}$) brings the wall velocity of the CS nanowire very close to the FM case. 
(b) FM and CS nanowires along [011].
$R_c=5a, \tsh=3a, L_c=50a, H=-0.5\Jfm$ and $T=0.01\Jfm$.
Straight lines are linear fit to the data.
}
\label{fig:zpos}
\end{figure}

The presence of the AF shell modifies the domain wall in the FM core as discussed next.
First, the domain wall width increases slightly ($\delta_{FM}\approx 7a$ and  $\delta_{CS} \approx 10a$) as can be seen by inspection of Fig.~\ref{fig:mprof}.
The domain wall width is determined from the competition between exchange and effective anisotropic energy, which also includes the shape-induced anisotropy contribution\cite{thi02},
$\delta=\pi \sqrt{A/(K+\pi M_s^2)}$.
In an atomistic description, an analogous expression would read
$\delta=a\pi\sqrt{J/(K+\lambda g \langle S_z \rangle^2)}$, 
where $\langle S_z \rangle$ is the total magnetization (average spin) of the composite  nanowire and $\lambda$ a geometrical constant. 
As seen in Fig.~\ref{fig:loops} the total magnetization of the AF-coated nanowire is lower than the FM one and this fact explains the increase of the domain wall width. 
Second, we observe an almost uniform lowering of the magnetization in the part of the nanowire and between the pair of domain walls, which is more pronounced in the case of a highly uncompensated interface (Fig.~\ref{fig:mprof_rot}) than in the weakly uncompensated one (Fig.~\ref{fig:mprof}). 
Since this drop of magnetization occurs far from the region of the domain walls, we interpret it as a secondary reversal mechanism.
The fact that the magnetization profile between the domain walls remains uniform while the values are reduced, points to a spatially confined coherent mechanism. 
Additionally, the fact that it is only observed in the core-shell nanowire implies that this mechanism must be related to the dynamics of the shell interface layer. 
To explain this mechanism, the following scenario is put forward. Owing to the strong interface coupling in our model ($|\Jint| \sim \Jfm, |\Jaf| $), the shell interface moments are dragged by the core interface moments and perform a reversible motion, which acts as a moving boundary to the core and promotes the coherent rotation of the magnetization in the region between the domain walls.
To confirm this scenario, we have performed a test simulation in which we have kept frozen the  shell interface moments during the magnetization reversal and by doing so, the magnetization profiles and wall velocities of the FM and the CS nanowire became almost identical, as discussed below (Fig.~\ref{fig:zpos}). 
This observation justifies the conjecture that the origin of the secondary reversal mechanism observed in the central region of the CS nanowire has its origin in the dynamics of the shell interface moments.

This is not the full picture yet, since in the case of the CS[011] nanowire (Fig.~\ref{fig:mprof_rot}) that has a highly uncompensated interface, freezing the shell moment has not eliminated completely the coherent-like mechanism and the uniform lowering of the magnetization in the central part of the nanowire has remained. 
To explain this distinct behavior of the CS[011] nanowire examined the detailed structure of the interface layer.
The interface shell moment of the CS[011] nanowire in the FC state,  is large and positive ($M_{shif} \approx 0.4$), that is, parallel to the core magnetization, despite the AF character of the interface coupling ($\Jint <0$). 
So the coherent rotation in the central part of core becomes a fast process as it lowers both the Zeeman energy and the interface exchange  energy of the unsatisfied FM-AF bonds. 

A similar conclusion regarding the role of the AF interface moments in a Co/CoO nanowire was drawn by Maurer \textit{et al}\cite{mau09}, who argued that the frozen moments of AF grains in the shell act as nucleation centers that promote the core magnetization reversal. 
In our atomistic model of a contiguous shell, the shell moments act collectively and the unsatisfied FM-AF bonds promote the reversal in a similar manner. Since they are uniformly distributed in the lateral surface of the core in the CS[011] nanowire, the resultant nucleation processes take the form of a coherent-like process in the central part of the wire. 

However, this argument cannot to explain the magnetization profile lowering in the case of the the CS system (Fig.~\ref{fig:mprof}b), because it has a very small number of unsatisfied bonds and very low shell interface moment ($M_{shif} \approx 0.01$) and thus only the reversible motion of the shell interface moments facilitates the reversal of the core, as explained previously. 
However, the zig-zag features of the central part of the profile (Fig.~\ref{fig:mprof}b) could be understood as a reminiscent of the coherent reversal mechanism and reflect the tendency of half of the shell interface spins (unsatisfied bonds) to act as nucleation centers for a coherent rotation mechanism. 

To quantify the characteristics of the wall dynamics, we consider next the wall velocity and mobility under application of a reversing field to a nanowire in the FC state. 
In the viscous regime of domain wall propagation, which we model here, the domain wall velocity is linear in the applied field, namely\cite{all08,ber98} 
\begin{equation}
  v(H)=\mu\cdot(H-H_{o}) 
  \label{eq:mob}
\end{equation}
where $\mu$ is the field-independent wall mobility, $H_o$ is the left coercive field (Fig.~\ref{fig:loops}) and the field values are taken $H > H_o$.
$  $
\begin{figure}[htb!]
	\centering
	\includegraphics[width=0.95\linewidth]{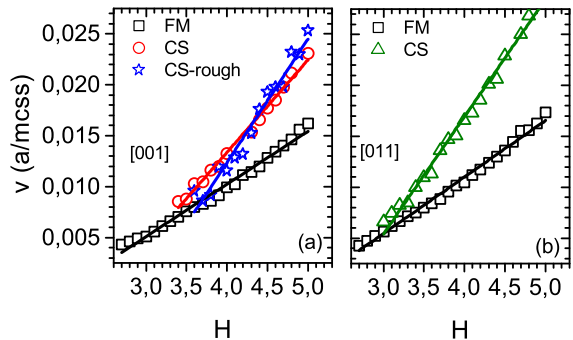}
	\caption{
		Field-dependence of domain wall velocity.
		(a) FM, CS and CS-rough nanowires along [001]. 
		(b) FM and CS nanowires along [011].
		$R_c=5a, \tsh=3a, L_c=50a, H=-0.5\Jfm$ and $T=0.01\Jfm$.
		Straight lines are linear fit to the data.
	} 	
	\label{fig:mob}
\end{figure}
The wall velocity $v(H)$ is obtained from a linear fit to the time evolution of the position of the  domain wall center, $z_o$, defined as the  point satisfying $M_z(z_o)=0$ in Fig.~\ref{fig:mprof} and Fig.~\ref{fig:mprof_rot}. 
Typical results for the wall displacement are shown in Fig.~\ref{fig:zpos}.
The deviations from linearity in the wall displacement seen at late time steps in Fig.~\ref{fig:zpos}, signify the beginning of domain wall annihilation.
This is more clearly seen in the CS nanowire as the wall widths are larger than the FM nanowire. 
In Fig.~\ref{fig:zpos}, we also show the decrease of wall velocity for  a CS nanowire with frozen shell. 
As discussed previously, the velocity reduction occurring when the shell moments are kept frozen signifies the role of dynamics of the shell interface moments in enhancing the domain wall velocity. 
The wall velocity at a certain applied field, Eq.~(\ref{eq:mob}), depends also on the intrinsic properties of the system which are contained in the physical parameter of mobility.
We calculate the wall mobility $\mu$ from a linear fit to the field-dependent velocity, as shown in Fig.~\ref{fig:mob}.
The domain wall velocity from Fig.~\ref{fig:zpos} and mobility from Fig.~\ref{fig:mob} are summarized in Table~\ref{table:dw}.
\begin{table}   
  \caption{ 
  Domain wall characteristics of nanowires with 
  $\Rc=5a, \tsh=3a$ and $\Lc=50a$ at $T=0.01\Jfm$
  }
  \begin{ruledtabular}
  \begin{tabular}{ l l l l }            
    System         &$v^{(1)}$  &$\mu^{(2)}$  &$\Heb$ \\
    \hline
    FM[001]        &0.016      &0.006        & - \\
    CS[001]        &0.021      &0.009        &0.15\\
    CS[001]-rough  &0.028      &0.012        &0.20\\
    FM[011]        &0.017      &0.006        & - \\
    CS[011]        &0.25       &0.012        &0.21\\
  \end{tabular}
  \end{ruledtabular}
  $^{(1)}$ Velocity values calculated at $H=-0.5~\Jfm$; 
  units are in $a/mcss$  \\
  $^{(2)}$ Mobility units are in $a/(mcss \cdot \Jfm)$ \\
  \label{table:dw}
\end{table}
These data show an increase of wall mobilty, due to coating of a FM nanowire by an AF shell. 
Additional increase of mobility is observed when the degree of uncompensation increases, which is achieved either by introducing interface roughness (rough-CS[001]) or by changing the crystallographic orientation of the FM-AF interface (CS[011]). 
Therefore, the uncompensated interface moments of the rough-CS[001] and the CS[011] nanowires drive the partially coherent reversal mechanism observed around the central part of the wire, which further facilitates the propagation of the wall pair. 

\section{Conclusions}
We performed atomistic modeling of the exchange bias effect in FM-core/AF-shell cylindrical nanowires using the Metropolis Monte Carlo algorithm. 
Our results showed that the interface exchange coupling causes reduction of the coercivity and appearance of a weak exchange bias effect, in reasonable agreement with recent measurements in Co/CoO nanowires\cite{mau09}.
The exchange bias effect is weaker in nanowires than in spherical core-shell nanoparticles due to the cylindrical shape of the interface region, that favors compensation of the AF moments along the wire axis.
Strong interface coupling induces a reversible motion of shell moments during reversal of the core magnetization, which provides  the dominant mechanism of enhanced domain wall mobility in nanowires with  highly compensated interfaces. 
As the degree of uncompensation, or equivalently the strength of the exchange bias field increases, for example due to roughness or disorientated wire axis, a secondary reversal mechanism is activated that has the characteristics of a coherent magnetization reversal, confined, though, in the central part of wire. 
This mechanism stems from the unsatisfied bonds that act as a uniform distribution of nucleation centers along the surface of the core. 
This secondary mechanism acts in synergy to the domain wall propagation leading to even further increase of the wall mobility observed in the system.
Finally, the technological implications of enhanced wall mobility in FM wires, due to exchange coupling to a thin oxide layer, could motivate  further investigation of this subject. 

\begin{acknowledgments}
DK acknowledges the hospitality in the Physics Department, University of Athens, during the course of this work and the financial support for dissemination of the work by the School of Pedagogical and Technological Education through project \textit{Strengthening of Research of ASPETE Faculty Members 2015-2016}.
\end{acknowledgments}


\begin{thebibliography}{38}
\bibitem[{com()}]{comm1b}
	\bibinfo{note}{For cobalt, we assumed discretized parameters on a SC lattice
		with cell size $\Delta=3nm$ and material parameters $A=1.3\cdot10^{-11}~J/m,
		Ms=1.4\cdot10^6~A/m$ and $K_u=3\cdot 10^5~J/m^3$. For cobalt oxide, we
		assumed for simplicity, the same net moment per cell, $A_{AF}=0.5A_{FM}$ and
		$K_{u,AF}=10 K_{u,FM}$.}
\end{thebibliography}
\end{document}